\documentclass[aps,superscriptaddress,twocolumn,twoside,floatfix,nofootinbib]{revtex4}
\usepackage{graphicx,amsmath,amssymb,color,array}
\usepackage[normalem]{ulem}
\usepackage{multirow}
\usepackage{times}

\usepackage{amsfonts}
\usepackage{latexsym}
\usepackage{amsmath}
\usepackage{amssymb}
\usepackage{amsfonts}
\usepackage{amsthm}
\usepackage{mathrsfs}
\usepackage{natbib}
\usepackage{color,verbatim,graphics}
\usepackage{psfrag}
\DeclareMathAlphabet{\mathrsfs}{U}{rsfs}{m}{n}
\DeclareMathAlphabet{\mathpzc}{OT1}{pzc}{m}{it}
\DeclareMathAlphabet{\matheus}{U}{eus}{m}{n}
\DeclareMathAlphabet{\mathbbold}{U}{bbold}{m}{n}

\newcommand{\ba}{\begin{eqnarray}}
\newcommand{\be}{\begin{equation}}
\newcommand{\ee}{\end{equation}}
\newcommand{\beq}{\begin{equation}}
\newcommand{\eeq}{  \end{equation}}
\newcommand{\bea}{\begin{eqnarray}}
\newcommand{\eea}{  \end{eqnarray}}

\newcommand{\ea}{\end{eqnarray}}
\newcommand{\ban}{\begin{eqnarray*}}
\newcommand{\ean}{\end{eqnarray*}}
\newcommand{\Tr}{\operatorname{tr}}

\newcommand{\ket}[1]{|#1\rangle}
\newcommand{\bra}[1]{\langle#1|}

\newcommand{\expect}[1]{\langle#1\rangle}
\newcommand{\assem}[3]{#1_{#2}^{#3}}
\newcommand{\tsum}{\textstyle{\sum}}

\newcommand{\eg}{{\it{e.g.}}}
\newcommand{\ie}{{\it{i.e.}~}}

\newcommand{\rA}{\mathrm{A}}
\newcommand{\rB}{\mathrm{B}}
\newcommand{\rC}{\mathrm{C}}
\newcommand{\rT}{\mathrm{T}}
\newcommand{\rGHZ}{\mathrm{GHZ}}
\newcommand{\rNS}{\mathrm{NS}}

\newcommand{\obs}{\mathrm{obs}}
\newcommand{\id}{\mathrm{id}}
\newcommand{\dani}[1]{{ #1}}

\usepackage{graphicx}
\begin{document}

\title{Detection of entanglement in asymmetric quantum networks and multipartite quantum steering}

\author{D. Cavalcanti}
\email{daniel.cavalcanti@icfo.es}
\author{P. Skrzypczyk}
\affiliation{ICFO-Institut de Ciencies Fotoniques, Mediterranean Technology Park, 08860 Castelldefels (Barcelona), Spain}
\author{G. H. Aguilar}
\author{R. V. Nery}
\author{P. H. Souto Ribeiro}
\author{S. P. Walborn}
\affiliation{Instituto de F\'{\i}sica, Universidade Federal do Rio de Janeiro, CP 68528, 21941-972, Rio de Janeiro, RJ, Brazil}

\begin{abstract}
The future of quantum communication relies on quantum networks composed by observers
sharing multipartite quantum states. The certification of multipartite entanglement will be crucial to the
usefulness of these networks. In many real situations it is natural to assume that
some observers are more trusted than others in the sense that they have more knowledge of their
measurement apparatuses. Here we propose a general method to certify all kinds of multipartite
entanglement in this asymmetric scenario and experimentally demonstrate it in an optical experiment. Our
results, which can be seen as a definition of genuine multipartite quantum steering, gives a method to detect entanglement in a scenario in-between the standard entanglement and fully device-independent scenarios, and provides a basis for semi-device-independent cryptographic applications in quantum networks.
\end{abstract}

\maketitle

\section{Introduction}

The most widely used techniques to detect entanglement rely either on having knowledge of the quantum state, obtained through quantum state tomography, or on the use of measurements that constitute an entanglement witness \cite{ent detection}. A frequently disregarded assumption behind these methods is that the measurements and devices used are well characterized. However, a mismatch between the theoretical description of the measurements and their actual implementation may lead to erroneous conclusions about the presence of entanglement \cite{Rosset}. A way of avoiding this assumption is to use device-independent techniques \cite{Bell review}, where the measuring devices are not trusted to behave as expected, and no specific description of the experimental observables is assumed. In this approach the measurement devices are considered as black boxes that the parties can access with classical inputs (corresponding to the measurement choices) that provide classical outputs (considered as the measurement results). The presence of entanglement is then verified analyzing the correlation statistics between the data lists corresponding to the measurement results. The violation of Bell inequalities \cite{Bell64} certify the presence of entanglement in this scenario, which can be thought of as a device-independent entanglement witness. The device-independent approach is especially important in adversarial scenarios, such as device-independent quantum key distribution \cite{DIQKD}, where an adversary can use a mismatch between the real implementation of the protocol and its description to fake its performance \cite{hack1,hack2,hack3}. However, the violation of a Bell inequality requires a high degree of correlation between the parties tolerating then very low levels of noise and demanding highly efficient detectors and high-quality entangled states \cite{Bell review}.

An intermediate scenario between the standard and the device-independent cases is that of quantum steering \cite{Schr,WJD07}. This is the situation where, in the bipartite case, one of the parties uses a trusted measuring device but the other does not. As such, we refer to this approach as the semi-device-independent one. Apart from the fundamental importance of characterizing separability in different scenarios, quantum steering appears as a practical situation that is less demanding experimentally than the device-independent approach. It requires fewer assumptions than the standard case and lower strength for the quantum correlations to be witnessed or certified. For these reasons the study of quantum steering, including its applications \cite{1sided,SteeringRand} and experimental demonstrations \cite{exp steering0, exp steering1, exp steering2,exp steering3,exp steering4,exp steering5,exp steering6}, have increased rapidly over recent years.

In the multipartite case, much knowledge has been acquired concerning standard entanglement detection \cite{ent detection} and the device-independent case \cite{di mult nl1,di mult nl2,di mult nl3,DIEW,expDIEW,McKague,SelfT1,SelfT2}. However, only few results were found in the semi-device-independent case. For instance, Ref.~\cite{CavMulti} provides inequalities to rule out fully separable states, Ref.~\cite{Pappa} developed a probabilistic protocol to detect the presence of a particular multipartite entangled state, and Ref.~\cite{Reid multi} discussed a hybrid model where each party is sometimes trusted and sometimes untrusted (see also \cite{Seiji, Li2015} for recent experimental demonstrations).

Apart from the fundamental problem of understanding multipartite quantum correlations, extending the semi-device-independent approach to the multipartite scenario is also relevant for practical purposes. As technology advances it will be possible to establish large quantum networks. These networks will be asymmetric in many cases, depending on the experimental capabilities of each station, the specific architecture of the implemented protocols, and unavoidable limitations that the setup may impose. Let us give few examples. Consider for instance prepare-and-measure cryptographic protocols in which some parties hold the sources of quantum systems and some others act as the receivers who measure these systems. Since the senders do not receive any external signals they may consider that no eavesdropper is manipulating their apparatuses. Thus, any error they observe (as for example due to detection inefficiencies) can be attributed to the apparatuses' imperfections. The receivers, on the other hand, are given systems that may have been intercepted by an eavesdropper, who may use extra degrees of freedom that are not considered by the receiver (see \cite{hack1,hack2,hack3} for examples). In this case, the receivers' apparatuses can not be considered trusted. Another scenario is that in which no reference frame can be established by some of the stations \cite{RefFrame}. In this case, the measurement directions that some of parties implement are not known, and may as well be considered as untrusted. Finally, quantum-key-distribution systems and quantum randomness generators are nowadays at the commercial level. Clearly, the general consumers of these products are not capable of reverse engineering the devices, and may not want to trust their providers.

Here we propose a general method to detect all kinds of entanglement that can be present in a quantum network, where some of the parties use untrusted measurements and must use data lists. We show how the different types of entanglement constrain the corresponding observed experimental data and present an efficient method to obtain semi-device-independent entanglement witnesses in the form of multipartite steering inequalities. We furthermore implement this method in a proof-of-principle optical experiment and demonstrate the violation of multipartite steering inequalities in both scenarios where either one or two parties perform untrusted measurements. Finally, we also quantify the advantage that the present approach provides over the device-independent one in terms of tolerance to noise.



\section{Results}
\subsection{Semi-device-independent test of multipartite entanglement}

\begin{figure}
  \includegraphics[width=8cm]{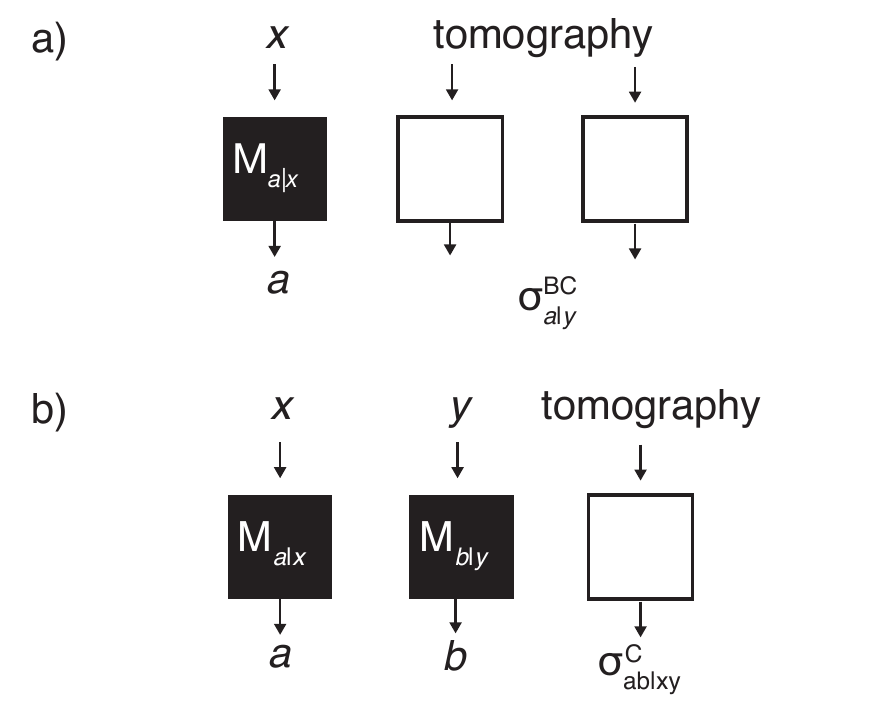}
\caption{\textbf{Asymmetric tripartite networks where untrusted devices are treated as black boxes with classical inputs and outputs} a) One untrusted party scenario: Alice, who holds an untrusted device treats it as a black box in which she inputs $x$ (the measurement choice) and receives an output $a$ (the measurement outcome). This procedure corresponds mathematically to applying some unknown measurement operator M$_{a|x}$ to the shared tripartite quantum state, which produces a post-measured state $\sigma^{\rB\rC}_{a|x}$ at Bob and Charlie's locations. \textbf{b)}  A similar situation occurs in two untrusted parties scenario, when both Alice and Bob perform untrusted measurements (corresponding to unkown measurement operators M$_{a|x}$ and M$_{b|y}$ respectively) preparing quantum states $\sigma^{\rC}_{ab|xy}$ on Charlie's system.}
\label{fig:scheme}
\end{figure}

We start by explaining the scenario considered here, which consists of a quantum network on $N$ parties sharing an unknown system in state $\rho$ (see Fig. \ref{fig:scheme}). Some of the parties perform measurements that are uncharacterised, or untrusted, while others have total control over their measurement apparatuses. Those parties who do not trust their apparatuses treat them as black boxes in which they can provide classical inputs (corresponding to the choice of measurement settings) and receive classical outputs (corresponding to the measurement outcomes). Notice that not even the Hilbert space dimension of these systems are assumed. The parties that trust their measurements can actually implement quantum state tomography, and reconstruct the density matrix they hold after the untrusted parties announce their measurement choices and outcomes. Based on this knowledge the goal is to decide if the original state $\rho$ had some kind of entanglement.

In the general case of $N$ parties there will be several semi-device-independent cases, depending on which parties are trusted. For simplicity, in what follows we will explain our method for the case of detecting genuine multipartite entanglement in a tripartite system. This case contains all the basic ingredients needed to understand both how to detect other types of entanglement and how to treat systems composed of more parties. These procedures are described in detail in the Supplementary Notes I-III.

Let us consider that an unknown tripartite state $\rho^{\rA\rB\rC}$ is distributed between three parties: Alice, Bob and Charlie.
Two semi-device-independent cases arise: (i) when only one party's device is untrusted and (ii) when two parties' devices are untrusted. Let us consider the first case, supposing that Alice holds the untrusted device. In this case, there is no assumptions on Alice's measurements and we describe them with some unknown measurement operators $M_{a|x}$, where the subscript $x$ labels the measurement choices and $a$ the possible outcomes. Not even the dimension of Alice's subsystem is assumed. Since Bob and Charlie trust their apparatuses they can perform tomography and determine their (unnormalized) conditional states $\assem{\sigma}{a|x}{\rB\rC}$ as
\be\label{untrusted A}
\assem{\sigma}{a|x}{\rB\rC}=\Tr_\rA(M_{a|x}\otimes \openone_{\rB}\otimes\openone_{\rC}\rho^{\rA\rB\rC}).
\ee
The set of unnormalized states $\{\assem{\sigma}{a|x}{\rB\rC}\}_{a,x}$ is called an assemblage and contains all the information obtainable in this situation as it encodes both the probability that Alice obtains the result $a$ given that she made the measurement $x$, as $p(a|x)=\Tr(\assem{\sigma}{a|x}{\rB\rC})$, as well as the corresponding conditional state $\assem{\rho}{a|x}{\rB\rC}=\assem{\sigma}{a|x}{\rB\rC}/p(a|x)$.

The second situation is when two parties, say Alice and Bob, have untrusted devices. In this situation Bob's measurement is also treated as a black box performing
measurements associated to unknown measurement operators $M_{b|y}$, while Charlie can tomographically determine the assemblage
\be\label{untrusted AB}
\assem{\sigma}{ab|xy}{\rC}=\Tr_{\rA\rB}(M_{a|x}\otimes M_{b|y}\otimes \openone_\rC\rho^{\rA\rB\rC}).
\ee
The probability distributions of Alice and Bob's measurements is encoded in $p(ab|xy)=\Tr\assem{\sigma}{ab|xy}{\rC}$.

If the initial state $\rho^{\rA\rB\rC}$ contains no genuine multipartite entanglement, \ie it is biseparable, then it has the form
\ba\label{bisep rho}
\rho^{\rA\rB\rC}&=&\sum_\lambda p_\lambda^{\rA:\rB\rC}\rho^\rA_\lambda\otimes\rho^{\rB\rC}_\lambda+\sum_\mu p_\mu^{\rB:\rA\rC}\rho^\rB_\mu\otimes\rho^{\rA\rC}_\mu\nonumber\\&+&\sum_\nu p_\nu^{\rA\rB:\rC}\rho^{\rA\rB}_\nu\otimes\rho^{C}_\nu,
\ea
where $p_\lambda^{\rA:\rB\rC}$, $p_\mu^{\rB:\rA\rC}$ and $ p_\nu^{\rA\rB:\rC}$ are probability distributions.  Then the assemblages \eqref{untrusted A} and \eqref{untrusted AB} have the form
\ba\label{gen mult steering two parties_mt}
\assem{\sigma}{a|x}{\rB\rC}&=&\Tr(M_{a|x}\otimes \openone_\rB\otimes\openone_\rC \rho^{\rA\rB\rC})\nonumber\\
&=&\sum_\lambda p_\lambda^{\rA:\rB\rC}p_\lambda(a|x)\rho^{\rB\rC}_\lambda \label{4a}\\
&+&\sum_\mu p_\mu^{\rB:\rA\rC}\rho^\rB_\mu\otimes\assem{\sigma}{a|x,\mu}{\rC}\label{4b}\\
&+&\sum_\nu p_\nu^{\rA\rB:\rC}\assem{\sigma}{a|x\nu}{\rB}\otimes\rho^{\rC}_\nu\label{4c}
\ea
and
\ba\label{gen mult steering single party_mt}
\assem{\sigma}{ab|xy}{\rC}&=&\Tr_{\rA\rB}(M_{a|x}\otimes M_{b|y}\otimes\openone_\rC \rho^{\rA\rB\rC})\nonumber\\
&=&\sum_\lambda p_\lambda^{\rA:\rB\rC}p_\lambda(a|x)\assem{\sigma}{b|y,\lambda}{\rC}\label{5a}\\
&+&\sum_\lambda p_\mu^{\rB:\rA\rC}p_\mu(b|y)\assem{\sigma}{a|x,\mu}{\rC}\label{5b}\\
&+&\sum_\lambda p_\nu^{\rA\rB:\rC}p_\nu(ab|xy)\rho^{\rC}_\nu\label{5c}
\ea
respectively.

Thus, the fact that the original state is biseparable imposes constraints on the observed assemblages. 
For instance, in \eqref{4a} the dependence on the variables $a$ and $x$ is only through the distribution $p_\lambda(a|x)$ and not through the quantum states. This is a typical instance of an unsteerable bipartite assemblage \cite{WJD07}. The assemblage in \eqref{4b} satisfies two constraints: each conditional state is a separable state, and the dependence in $a$ and $x$ is due only to Charlie's system, and not Bob's. The assemblage in \eqref{4c} is similar to the one in \eqref{4b}, only with Bob's and Charlie's roles exchanged. Thus, in order to test if a given assemblage $\assem{\sigma}{a|x}{obs}$ has the form \eqref{gen mult steering two parties_mt} consistent with having been produced by a biseparable state one could run the following program:
\ba\label{gen mult steering two parties_mt2}
&&\text{find}~\assem{\Gamma}{a|x}{\rA:\rB\rC},\assem{\Gamma}{a|x}{\rB:\rA\rC},\assem{\Gamma}{a|x}{\rC:\rA\rB}, \\
&&\text{such that}\nonumber\\
&&\assem{\sigma}{a|x}{obs}=\assem{\Gamma}{a|x}{\rA:\rB\rC}+\assem{\Gamma}{a|x}{\rB:\rA\rC}+\assem{\Gamma}{a|x}{\rC:\rA\rB}, \nonumber\\
&&\assem{\Gamma}{a|x}{\rA:\rB\rC}\geq0,~\assem{\Gamma}{a|x}{\rB:\rA\rC}\geq0, \assem{\Gamma}{a|x}{\rC:\rA\rB}\geq0, \nonumber\\
&&\assem{\Gamma}{a|x}{\rA:\rB\rC} \text{is unsteerable},\nonumber\\
&&\assem{\Gamma}{a|x}{\rB:\rA\rC} \text{is separable and unsteerable from $A$ to $B$},\nonumber\\
&&\assem{\Gamma}{a|x}{\rC:\rA\rB} \text{is separable and unsteerable from $A$ to $C$}.\nonumber
\ea
If no such triple of assemblages exists, then the underlying state was definitely not biseparable, and therefore genuine multipartite entangled. The main problem with this method is that, apart from systems with dimension lower than $6$, testing separability is computationally demanding  \cite{gurvitz}. As we show in the Supplementary Notes I and II, we can overcome this problem by considering approximations of the set of separable states which relax the above program into a semidefinite program (SDP) \cite{DPS,DPSmulti}, for which efficient numerical methods exist.

A similar analysis can be made for the decomposition \eqref{gen mult steering single party_mt} (see Supplementary Note II) and other types of entanglement (see Supplementary Table I). For instance, \eqref{5a} refers to an assemblage that is unsteerable from $A$ to $C$ and \eqref{5b} to one that is unsteerable from $B$ to $C$. The assemblage \eqref{5c} has two properties: it is unsteerable, and the probability distributions $p_\nu(ab|xy)$ must have quantum realisations, \ie must come from measurements on quantum states. Again, this last requirement is in general difficult to test. However, we can once again make use of relaxations of the set of quantum probability distributions \cite{NPA} to transform the program into a semidefinite program.

All in all, for each semi-device-independent scenario the type of entanglement in the distributed state will impose constraints on the assemblages one could observe. These constraints allow the parties holding the trusted devices to determine if this state must have been genuine multipartite entangled (\eg if the observed data admits no decomposition of the form \eqref{gen mult steering two parties_mt} or  \eqref{gen mult steering single party_mt} then there exists no biseparable state that could explain it). Therefore, even not knowing the initial state or what type of measurements the untrusted parties performed, it is possible to discriminate the assemblages that were produced by states containing some type of entanglement.

Finally, in each case the program can be seen as a membership test for the observed assemblage to be contained inside a convex set. It is always possible to certify that a point lies outside a convex set by finding a separating hyperplane between the set and the point. As we show in the Supplementary Note II, in each case we can find the lagrange dual program to the set membership test, which always amounts to finding such a separating hyperplane. Such separating hyperplanes are precisely multipartite steering inequalities, which can alternatively be thought of as semi-device independent entanglement witnesses. Thus our method naturally generates steering inequalities which can then be used as witnesses for multipartite entanglement.

\subsection{Practical considerations}\label{sec:Practical}

Due to experimental errors and finite statistics the experimentally observed data is not strictly compatible with any physical state and local measurements. In particular, all assemblages which exactly reproduce the experimental data in general do not satisfy the no-signalling constraint that $\sum_a \assem{\sigma}{a|x}{\rB\rC}=\sum_a\assem{\sigma}{a|x'}{\rB\rC}$ for $x\neq x'$.
Since the present methods are tailored to detect entanglement of physical states we can not use the observed data directly to test for the presence of entanglement.

We thus propose to proceed with the following steps: First, given the experimental data, generate a physical assemblage that best approximates it through, for instance, a maximum likelihood reconstruction method. Second, having obtained the best physical approximation to the actual data, use the SDP method discussed in the Supplementary Note II to check for any type of entanglement. This method also generates an inequality that is satisfied by all assemblages coming from states which do not have the type of entanglement tested. Finally check that the observed data violates this inequality.

\subsection{Examples: GHZ and W states}
As examples, we used our method to produce the following inequality that is satisfied by all assemblages of the form \eqref{untrusted A} (see also Supplementary IV):
\begin{multline}\label{eq:GHZoneuntrusted}
1
+ 0.1547 \expect{Z_\rB Z_\rC}
- \tfrac{1}{3}\big(\expect{A_3 Z_\rB} +\expect{A_3 Z_\rC} + \expect{A_1 X_\rB X_\rC} \\- \expect{A_1 Y_\rB Y_\rC} - \expect{A_2 X_\rB Y_\rC} - \expect{A_2 Y_\rB X_\rC}\big) \geq 0,
\end{multline}
with $A_i$ for $i = 1, 2, 3$, being observables in Alice's system with outcomes labelled $\pm1$ and $X, Y$ and $Z$ representing the Pauli operators. The GHZ state $(\ket{000}+\ket{111})/\sqrt{2}$ violates this inequality by $-0.8453\ngeq 0$ when Alice's measurements are also $X$, $Y$ and $Z$, which numerical optimization suggests are the optimal choices for Alice.

In the case Alice and Bob perform untrusted measurements we have derived the following inequality which is satisfied by assemblages of the form \eqref{untrusted AB}:
\begin{multline}\label{eq:GHZtwountrusted}
1 - \alpha \langle A_3 B_3 \rangle - \alpha \langle A_3 Z \rangle - \alpha \langle B_3 Z \rangle  - \beta \langle A_1 B_1 X \rangle  \\
+ \beta \langle A_1 B_2 Y \rangle + \beta \langle A_2 B_1 Y \rangle + \beta \langle A_2 B_2 X \rangle \geq 0
\end{multline}
where $\alpha = 0.1831$ and $\beta = 0.2582$, and similarly $B_i$ for $i = 1, 2, 3$ represent Bob's measurement which we assume to have $\pm1$ outcomes. The GHZ state achieves a violation $-0.5820\ngeq 0$ now when both Alice and Bob perform $X$, $Y$ and $Z$ measurements.

Similar inequalities for two untrusted parties and for the W state, given by $\ket{W}=(\ket{001}+\ket{010}+\ket{100})/\sqrt{3}$, are presented in the Supplementary Note IV.

We have also considered noisy versions of the GHZ and W states given by
\ba
&\rho_{\psi}&=w\ket{\psi}\bra{\psi}+(1-w)\openone/8,
\ea
where $\ket{\psi}$ can be either the GHZ or the W state. We computed how much white noise can be added to these states until we are unable to detect genuine multipartite entanglement. Specifically, we quantify the minimum $w$ for which our method guarantees that following states are genuinely multipartite entangled. The results are summarized in Table \ref{examples}, together with the known bounds for standard entanglement tests \cite{GuhSee10,JunMorGuh12} and the device-independent case \cite{DIEW}.
One can see that trusting some of the parties offers a significant advantage in terms of noise tolerance.

\subsection{Experimental violation of genuine tripartite steering witnesses}

\begin{figure*}
\centering
  \includegraphics[width=12cm,keepaspectratio]{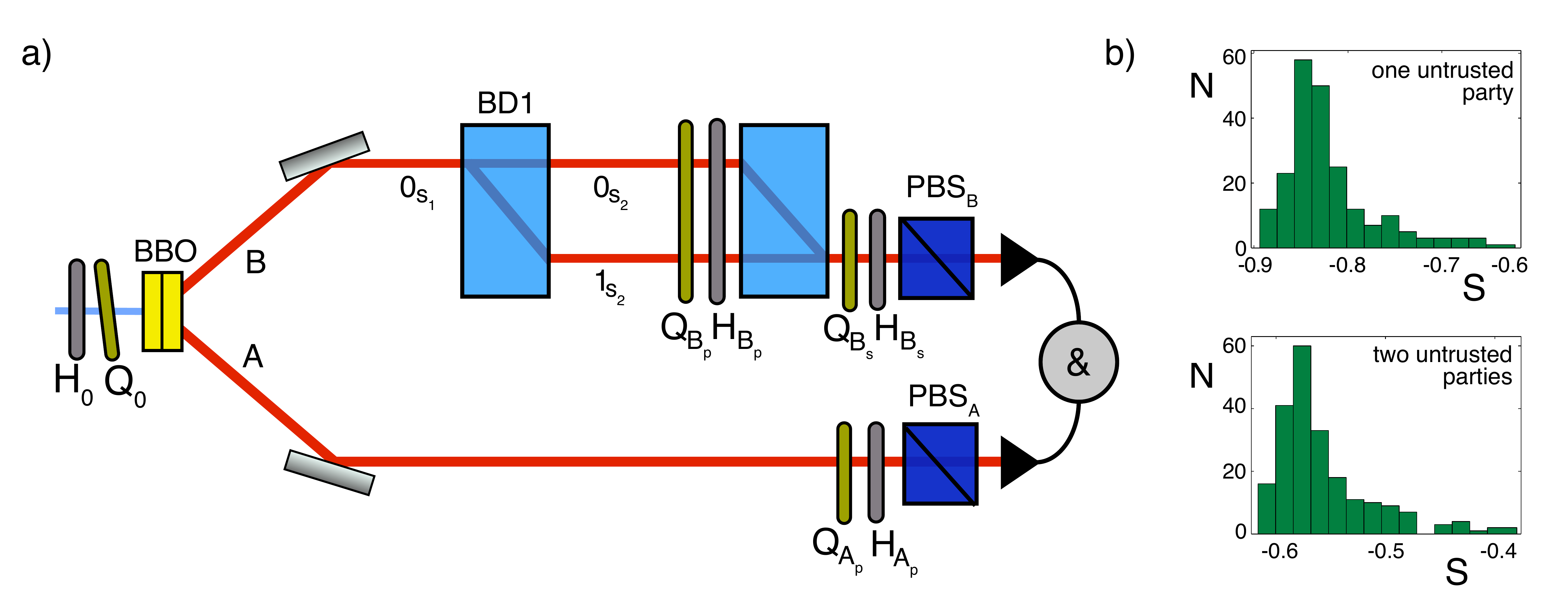}
\caption{\textbf{Experimental setup and results} 
a) An ultraviolet laser oscillating at 325nm pumps two cross axis 
BBO crystals, each 1mm long. Probabilistically, two photons are produced in the state \eqref{init_state} via parametric down 
conversion \cite{kwiat99}. The polarization entangled state is obtained by the superposition of signal and
idler beams produced in the first crystal with vertical polarization and the ones produced in the
second crystal with horizontal polarization. Signal photons in B are sent to beam displacer BD1 
which transmits the vertical polarization component and deviates the horizontally polarized component. 
This results in the production of a GHZ state right after BD1. Two qubits in this state are represented by the polarization degrees of freedom of photons A and B,
and one qubit is encoded in the spatial (or path) degree of freedom of photon B. 
Photons in mode A are sent directly to detection after polarization projection,
which is done using the quarter waveplate $\text{Q}_{\text{A}_\text{p}}$, half waveplate H$_{\text{A}_\text{p}}$,
and polarizing beam splitter PBS$_\text{A}$. We perform a joint analysis of the polarization and path degree of freedom of photon B
using the sequence of devices given by quarter waveplate Q$_{\text{B}_\text{p}}$, half waveplate H$_{\text{B}_\text{p}}$, 
beam displacer BD2, quarter waveplate Q$_{\text{B}_\text{s}}$, half waveplate H$_{\text{B}_\text{s}}$, and polarizing beam spliter
PBS$_\text{B}$. For a given set of adjustments of the quarter and half waveplates, we perform one specific
joint projection in both polarization and path degrees of freedom basis. 
Since there is a coherent combination of spatial modes 0 and 1 in BD2, the measure
of the spatial degrees of freedom of photon B is done by mapping the
spatial qubit state before BD2 in the polarization state at the output of BD2. 
Even though the projection is made simultaneously for both degrees of freedom in this case, 
they are independent, or in other words, all combinations of projections are possible \cite{farias12b, aguilar2014}. 
b) Histograms obtained by computing the semi-device-independent entanglement witness from the experimental data (see main text and the Supplementary Note V for more details about the witness). We measured the value of each witness 215 independent times and plot the number of times $N$ the value $S$ of the witness felt within a given interval. From this statistical procedure, an average value and a variance is obtained for the witness. The upper histogram is for the case of one untrusted party, resulting in the average value of -0.82 and standard deviation of 0.05. The lower histogram is for the case of two untrusted parties, resulting in the average of -0.56 and an standard deviation of 0.04.}
\label{fig:setup_mt}
\end{figure*}

In order to illustrate the utility and efficiency of our approach, we use this technique to violate genuine multipartite steering witnesses in a real laboratory setting where one or two parties perform untrusted measurements. The experimental set-up is shown in Fig \ref{fig:setup_mt} and is set to produce a GHZ state encoded in the polarization and path degree of freedom of two photons \cite{farias12b, aguilar2014} with high fidelity. The experimental procedure starts preparing photons in a state close to
\be
\ket{\Psi}=\left(\frac{\ket{00}+\ket{11}}{\sqrt{2}}\right)_{{\text{A}_p}{\text{B}_p}} \otimes\ket{0}_{\text{B}_s},
\label{init_state}
\ee
where $\text{A}_{p}$ and $\text{B}_{p}$ represent the polarization qubit of photons A and B respectively, where 0 and 1 stand for horizontal and vertical polarization states, and $\text{B}_{s}$ represents the spatial degree of freedom of photon B. To obtain a GHZ state, we couple the spatial degree of freedom with the polarization using Beam Displacer (BD1) which transforms  $\ket{0}_{\text{B}_{p}}\ket{0}_{\text{B}_{s}} \rightarrow\ket{0}_{\text{B}_{p}}\ket{0}_{\text{B}_{s}}$ and $\ket{1}_{\text{B}_{p}}\ket{0}_{\text{B}_{s}} \rightarrow \ket{1}_{\text{B}_{p}}\ket{1}_{\text{B}_{s}}$. Once we obtain the desired state, every qubit is measured in the eigenstates of the three Pauli operators. For the polarization degrees of freedom this is carried out using a quarter-wave-plate (QWP), a half-wave plate (HWP) and a polarizing beam splitter (PBS) or BD2, depending on the photon. For the spatial degrees of freedom this is carried out using the interferometer described in Fig. \ref{fig:setup_mt} \cite{farias12b, aguilar2014}.

Although this experiment is tailored to produce a GHZ state and perform measurements corresponding to the Pauli operators, the analysis we perform on the experimental data makes no assumption about the state nor the untrusted measurements. We consider two cases, one where part $\text{A}_{p}$ is untrusted and parts $\text{B}_{p}$ and $\text{B}_{s}$ hold the trusted devices, and another when parts $\text{B}_{p}$ and $\text{B}_{s}$ hold the untrusted devices and part $\text{A}_{p}$ the trusted one. For the two cases we follow the procedure described in Sec. \ref{sec:Practical} (using a least-squares optimisation to provide physical assemblages), which provides inequalities of the form $S\geq0$ (whose exact form can be found in the Supplementary Note V) whose violation certify that the corresponding assemblages cannot be written in the biseparable form \eqref{gen mult steering two parties_mt} or \eqref{gen mult steering single party_mt}, respectively. We finally observe a violation of these inequalities by the experimental data (see Fig. \ref{fig:setup_mt} b)). \dani{We have performed each experiment (\ie measuring all correlators) 215 independent times, from which we calculate an average value of $S=-0.82 \pm 0.05$ for one untrusted party and $S=-0.56 \pm 0.04$ for two untrusted parties.} This proves that there exists no biseparable tripartite state and measurements performed by the untrusted parties that could have generated the observed assemblages. 

\section{Discussion}

We have derived a method to detect multipartite entanglement when some of the apparatuses used in a quantum network are untrusted or uncharacterized. This method allows the detection of of all kinds of entanglement in quantum networks where some of the observers use their measurement apparatuses simply as data lists. This scenario is experimentally less demanding than the nonlocality scenario, as it tolerates more noise, for instance. We have performed a proof-of-principle experiment demonstrating the existence of genuine tripartite entanglement, without any assumption on the source or the measurements being performed in some of the subsystems.

Our results provide a feasible test for multipartite entanglement in quantum networks
and bridges the two well known cases of multipartite entanglement and multipartite Bell nonlocality. Moreover the scenario considered is a natural generalization of bipartite quantum steering \cite{WJD07} (see \cite{CavMulti,Reid multi} for alternative definitions) . Since steering has found applications in cryptographic protocols \cite{1sided,SteeringRand}, we believe that our results can be used as a starting point to define semi-device-independent cryptographic applications in future quantum networks.

\section{Acknowledgements}
We thank A. Ac\'in, M. Navascu\'es and L. Aolita for discussions. DC was supported by the Beatriu de Pin\'os fellowship (BP-DGR 2013) and PS by the Marie Curie COFUND action through the ICFOnest program. Financial support was provided by Brazilian agencies FAPERJ, CNPq,  CAPES, and the National Institute of Science and Technology for Quantum Information.


\begin{table}
\begin{center}
\begin{tabular}{c||c||c|c||c}
 {no. untr. meas.} & 0 & 1&2 &3 \\\hline\hline

 GHZ& $\frac{3}{7}\approx0.429$& $\approx0.54$ & $\approx0.63$ & $2/3\approx0.67$\\\hline\hline

 W &  $\approx0.479 $&$\approx0.57$  & $\approx0.67$ & $\approx0.72$
\end{tabular}
\end{center}
\caption{Critical robustness to white noise $w$.  We provide a comparison between the \dani{known bounds} on critical robustness to white noise of the GHZ and W states above which genuine multipartite entanglement can be detected in 4 different scenarios: when no party is untrusted (\ie the standard entanglement scenario \cite{GuhSee10,JunMorGuh12} ), when 1 and 2 parties hold untrusted devices, for which we used the semi-device-independent method developed here and when all devices are untrusted, \ie the device-independent case developed in \cite{DIEW}. In the Supplementary Table II we also display the bounds concerning the detection of (not necesarily genuine multipartite) entanglement in these states.}\label{examples}
\end{table}%

\begin{appendix}

\begingroup
\squeezetable
\begin{table*}
\begin{center}
\begin{tabular}{|m{3.2cm}|m{2cm}|m{2cm}|m{8cm}|}
\hline
Form of state & Unstrusted parties & Known objects & SDP \\
\hline
$$\sum_\lambda p_\lambda \rho_\lambda^\rA\otimes\rho_\lambda^\rB\otimes\rho_\lambda^\rC$$ & $$\rA$$ &$$\assem{\sigma}{a|x}{\rB\rC}$$&
\vspace*{-4mm}\vbox{\begin{align}
\text{max}&\quad p\nonumber\\
\text{s.t}&\quad  \tsum_\mu D_\mu(a|x)\assem{\sigma}{\mu}{\rB\rC} = \assem{\sigma}{a|x}{\rB\rC} - p\,\assem{\id}{a|x}{\rB\rC}, \\
&\quad  \left(\assem{\sigma}{\mu}{\rB\rC}\right)^{T_\rB}\geq 0,\quad \assem{\sigma}{\mu}{\rB\rC}\geq 0. \nonumber
\end{align} \vspace*{-6mm}}\\
\cline{2-4} & $$\rA\text{ and }\rB$$ & $$\assem{\sigma}{ab|xy}{\rC}$$ &
\vspace*{-4mm}\vbox{\begin{align}
\text{max}&\quad p\nonumber\\
\text{s.t.}&\quad\tsum_{\mu,\lambda} D_\mu(a|x)D_\lambda(b|y)\assem{\sigma}{\mu\lambda}{\rC} = \assem{\sigma}{ab|xy}{\rC} - p\,\assem{\id}{ab|xy}{\rC},\\
&\quad \assem{\sigma}{\mu\lambda}{\rC} \geq 0. \nonumber
\end{align}\vspace*{-6mm}}\\
\hline
$$\sum_{\lambda}p_\lambda\rho^\rA_\lambda\otimes\rho^{\rB\rC}_\lambda$$ & $$\rA$$ & \vbox{$$\assem{\sigma}{a|x}{\rB\rC}$$}&
\vspace*{-4mm}\vbox{\begin{align}\label{const1}
\text{max}& \quad p \nonumber \\
\text{s.t}& \quad \tsum_\mu D_\mu(a|x)\assem{\sigma}{\mu}{\rB\rC} = \assem{\sigma}{a|x}{\rB\rC} - p \, \assem{\id}{a|x}{\rB\rC}, \\
&\quad\assem{\sigma}{\mu}{\rB\rC} \geq 0. \nonumber
\end{align}\vspace*{-6mm} } \\
\cline{2-4} & $$\rB$$ & \vbox{$$\assem{\sigma}{b|y}{\rA\rC}$$} &
\vspace*{-4mm}\vbox{\begin{align}\label{const2}
\text{max}&\quad p \nonumber \\
\text{s.t}& \quad \assem{\Gamma}{b|y}{\rA\rC} = \assem{\sigma}{b|y}{\rA\rC} - p\, \assem{\id}{b|y}{\rA\rC}, \\
&\quad\Tr_\rC\assem{\Gamma}{b|y}{\rA\rC}=\tsum_\mu D_\mu(b|y)\assem{\sigma}{\mu}{\rA}, \nonumber\\
&\quad(\assem{\Gamma}{b|y}{\rA\rC})^{T_\rA}\geq 0, \quad \assem{\Gamma}{b|y}{\rA\rC} \geq 0, \quad \assem{\sigma}{\mu}{\rA} \geq 0.\nonumber
\end{align}\vspace*{-6mm} }\\
\cline{2-4} & $$\rA \text{ and } \rB$$ & $$\assem{\sigma}{ab|xy}{\rC}$$ &
\vspace*{-4mm}\vbox{\begin{align}\label{const3}
   \text{max}&\quad p\nonumber\\
   \text{s.t}&\quad\tsum_\mu D_\mu(a|x)\assem{\sigma}{b|y,\mu}{\rC} = \assem{\sigma}{ab|xy}{\rC} - p\, \assem{\id}{ab|xy}{\rC}, \\
& \quad  \assem{\sigma}{b|y,\mu}{\rC} \geq 0.\nonumber
\end{align}\vspace*{-6mm}}\\
\cline{2-4} & $$\rB\text{ and }\rC$$ & $$\assem{\sigma}{bc|yz}{\rA}$$ &
\vspace*{-4mm}\vbox{\begin{align}\label{const4}
\text{max}& \quad p\nonumber \\
\text{s.t.}&\quad\tsum_\mu D^{\rNS}(bc|yz,\mu)\assem{\sigma}{\mu}{\rA} = \assem{\sigma}{bc|yz}{\rA}-p\,\assem{\id}{bc|yz}{\rA}\\
&\quad\tsum_\mu D^{\rNS}(bc|yz,\mu)\assem{\sigma}{\mu}{\rA} \in \mathcal{Q}^{(k)}_{\rA},\quad \assem{\sigma}{\mu}{\rA} \geq 0 \nonumber
\end{align}\vspace*{-6mm}}\\
\hline
\vbox{\ba \sum_\lambda p_\lambda^{\rA:\rB\rC}\rho^\rA_\lambda\otimes\rho^{\rB\rC}_\lambda\nonumber\\+\sum_\lambda p_\lambda^{\rB:\rA\rC}\rho^\rB_\lambda\otimes\rho^{\rA\rC}_\lambda\nonumber\\+\sum_\lambda p_\lambda^{\rA\rB:\rC}\rho^{\rA\rB}_\lambda\otimes\rho^{C}_\lambda\nonumber \ea} & $$\rA$$ & $$\assem{\sigma}{a|x}{\rB\rC}$$&
\vspace*{-4mm}\vbox{\begin{align}
\text{max}&\quad p\nonumber\\
\text{s.t.}&\quad \assem{\Gamma}{a|x}{\rA:\rB\rC} + \assem{\Gamma}{a|x}{\rB:\rA\rC} + \assem{\Gamma}{a|x}{\rC:\rA\rB} = \assem{\sigma}{a|x}{\rB\rC} - p \,\assem{\mathrm{id}}{a|x}{\rB\rC}\nonumber\\
&\quad\assem{\Gamma}{a|x}{\rA:\rB\rC} = \tsum_\mu D_\mu(a|x)\assem{\sigma}{\mu}{\rB\rC}, \quad\assem{\sigma}{\mu}{\rB\rC} \geq 0\nonumber \\
&\quad \Tr_\rC \assem{\Gamma}{a|x}{\rB:\rA\rC} = \tsum_\mu D_\mu(a|x)\assem{\sigma}{\mu}{\rB},\quad \assem{\sigma}{\mu}{\rB} \geq 0,\\
&\quad \Tr_\rB \assem{\Gamma}{a|x}{\rC:\rA\rB} = \tsum_\mu D_\mu(a|x)\assem{\sigma}{\mu}{\rC},\quad \assem{\sigma}{\mu}{\rC} \geq 0,\nonumber\\
&\quad \left(\assem{\Gamma}{a|x}{\rB:\rA\rC}\right)^{T_\rB}\geq 0, \quad \left(\assem{\Gamma}{a|x}{\rC:\rA\rB}\right)^{T_\rB}\geq 0. \nonumber \\
&\quad\assem{\Gamma}{a|x}{\rB:\rA\rC}\geq 0, \quad \assem{\Gamma}{a|x}{\rC:\rA\rB}\geq 0, \quad\sum_{a} \assem{\Gamma}{a|x}{\rB:\rA\rC} = \sum_a \assem{\Gamma}{a|x'}{\rB:\rA\rC}.  \nonumber
\end{align}\vspace*{-6mm}}\\
\cline{2-4} & $$\rA\text{ and }\rB$$ & $$\assem{\sigma}{ab|xy}{\rC}$$&
\vspace*{-4mm}\vbox{\begin{align}
\text{max}&\quad p\nonumber\\
\text{s.t.}&\quad\assem{\Pi}{ab|xy}{\rA:\rB\rC} + \assem{\Pi}{ab|xy}{\rB:\rA\rC} + \assem{\Pi}{ab|xy}{\rC:\rA\rB} = \assem{\sigma}{ab|xy}{\rC} - p \,\assem{\mathrm{id}}{ab|xy}{\rC} \nonumber\\
&\quad \assem{\Pi}{ab|xy}{\rA:\rB\rC} = \tsum_\mu D_\mu(a|x)\assem{\sigma}{b|y,\mu}{\rC},\quad \assem{\sigma}{b|y,\mu}{\rC} \geq 0\nonumber \\
&\quad \assem{\Pi}{ab|xy}{\rB:\rA\rC} = \tsum_\mu D_\mu(b|y)\assem{\sigma}{a|x,\mu}{\rC},\quad \assem{\sigma}{a|x,\mu}{\rC} \geq 0 \\
&\quad\assem{\Pi}{ab|xy}{\rC:\rA\rB} = \tsum_\mu D_\nu^\rNS(ab|xy)\assem{\sigma}{\nu}{\rC},\quad \assem{\sigma}{\nu}{\rC} \geq 0\nonumber \\
&\quad \assem{\Pi}{ab|xy}{\rC:\rA\rB} \in \mathcal{Q}_\rC^{(k)},\quad\sum_b\assem{\sigma}{b|y,\mu}{\rC} = \sum_b\assem{\sigma}{b|y',\mu}{\rC} \nonumber
\end{align}\vspace*{-6mm}}\\\hline
\end{tabular}
\end{center}
\caption{\textbf{Collection of all SDP tests for the tripartite case.} All expressions with indices should be understood to hold for each value of the index. Various SDPs depend upon a parameter $k$, such that for larger values of $k$ we obtain a better approximate characterisation of the set, and therefore a more stringent test. All programs are strictly feasible and are such that a negative optimal value $p^* < 0$ certifies that the assemblage has the corresponding type of entanglement. In this case the dual provides a semi-device-indepenent entanglement witness in the form of a steering inequality. An optimal solution $p^* \geq 0$ indicates that the assemblage is inside the corresponding set, i.e. that one cannot conclude that the state contains the desired type of entanglement. }\label{t: SDPs}
\label{default}
\end{table*}%
\endgroup

\newpage
\begingroup
\squeezetable
\begin{table}
\begin{tabular}{ll||c||c|c||c|c||cc}\label{examples}
&   & EW  & \multicolumn{2}{c||}{SW} & \multicolumn{2}{c||}{SW} & \multicolumn{2}{c}{DIEW} \\
\multicolumn{2}{l||}{no. untr.} & 0 & \multicolumn{2}{c||}{1} & \multicolumn{2}{c||}{2} & \multicolumn{2}{c}{3} \\
\multicolumn{2}{l||}{no. meas.} & n/a & 2 & 3 & 2 & 3 & \multicolumn{1}{c|}{2} & 3 \\
\hline\hline
\multirow{2}{*}{GHZ} & ent. & \begin{tabular}[c]{@{}c@{}}$\tfrac{1}{5}$\\ = 0.2 \\ \cite{DurCir00}\end{tabular} & \begin{tabular}[c]{@{}c@{}}$\approx$\\$0.2613$\end{tabular} & \begin{tabular}[c]{@{}c@{}}$\approx$\\$0.2500$\end{tabular} & \begin{tabular}[c]{@{}c@{}}$\tfrac{1}{2}=$\\$0.5$\end{tabular} & \begin{tabular}[c]{@{}c@{}}$\tfrac{3}{7}\approx$\\$0.4286$\end{tabular} & \multicolumn{1}{c|}{ \begin{tabular}[c]{@{}c@{}}$\tfrac{1}{2}=$\\$0.5$ \\ \cite{GruLasZuk10}\end{tabular}} & \multicolumn{1}{c}{ \begin{tabular}[c]{@{}c@{}}$\tfrac{1}{2}=$\\$0.5$ \\ \cite{GruLasZuk10}\end{tabular}}  \\
\cline{2-9}
& GME  & \begin{tabular}[c]{@{}c@{}}$\tfrac{3}{7} \approx$\\ 0.4286 \\ \cite{GuhSee10}\end{tabular} & \begin{tabular}[c]{@{}c@{}}$\approx$ \\ $0.6307$\end{tabular} & \begin{tabular}[c]{@{}c@{}}$\approx$\\$0.5420$\end{tabular} & \begin{tabular}[c]{@{}c@{}}$\tfrac{1}{\sqrt{2}} \approx$\\ $0.7071$\end{tabular} & \begin{tabular}[c]{@{}c@{}}$\approx$\\$0.6322$\end{tabular}  & \multicolumn{1}{c|}{\begin{tabular}[c]{@{}c@{}}$\tfrac{1}{\sqrt{2}} \approx $\\ $0.7071$ \\ \cite{DIEW}\end{tabular}} & \begin{tabular}[c]{@{}c@{}}$\tfrac{2}{3} \approx$\\$0.6667$ \\ \cite{DIEW}\end{tabular} \\
\hline\hline
\multirow{2}{*}{W} & ent. & \begin{tabular}[c]{@{}c@{}}$\approx$\\0.2096 \\ \cite{Sza11}\end{tabular} & \begin{tabular}[c]{@{}c@{}}$\tfrac{3}{11}\approx$ \\ $0.2727$ \end{tabular} & \begin{tabular}[c]{@{}c@{}}$\approx$ \\ $0.2698$\end{tabular} & \begin{tabular}[c]{@{}c@{}}$\approx$ \\ $0.5765$\end{tabular} & \begin{tabular}[c]{@{}c@{}}$\approx$ \\ $0.4434$\end{tabular} & \multicolumn{1}{c|}{\begin{tabular}[c]{@{}c@{}}$ \approx $\\ $0.6442$ \\ \cite{GruLasZuk10}\end{tabular}} & \begin{tabular}[c]{@{}c@{}}$ \approx$\\$0.6048$ \\ \cite{GruLasZuk10} \end{tabular} \\
\cline{2-9}
& GME & \begin{tabular}[c]{@{}c@{}}$\approx$\\0.479 \\ \cite{JunMorGuh12}\end{tabular} & \begin{tabular}[c]{@{}c@{}}$\approx$\\$0.6440$\end{tabular} & \begin{tabular}[c]{@{}c@{}}$\approx$\\$0.5684$\end{tabular} & \begin{tabular}[c]{@{}c@{}}$\approx$\\$0.7218$\end{tabular} & \begin{tabular}[c]{@{}c@{}}$\approx$\\$0.6757$\end{tabular} & \multicolumn{1}{c|}{\begin{tabular}[c]{@{}c@{}}3/4\\ = 0.75 \\ \cite{DIEW}\end{tabular}} & \begin{tabular}[c]{@{}c@{}}$\approx$ \\ $0.7158$ \\ \cite{DIEW} \end{tabular}
\end{tabular}
\caption{\textbf{Critical robustness to white noise $w^*$ for different scenarios.} We provide a comparison between the critical robustness to white noise of the GHZ and W state, in the 4 different scenarios: All 3 parties with trusted devices, \ie using an entanglement witness (EW), 2 or 1 parties with trusted devices, \ie using a steering witness (SW) and no parties with trusted devices \ie nonlocality (DIEW). We give critical values for both detecting entanglement and for detecting genuine multipartite entanglement ($\approx$ refers to numerical results). }\label{examples}
\end{table}
\endgroup

\textbf{Supplementary note 1}\\

\textbf{Characterising multipartite assemblages}\\

\label{characterisation of assemblages}

Here we determine the constraints that different types of entanglement in the initial state $\rho^{\rA\rB\rC}$ impose on the assemblages produced by untrusted measurements and define the corresponding sets that they characterise. We will first consider the question of whether or not there is any entanglement in the state, before moving on to entanglement in a given bipartition, for which there are a number of different cases, given the asymmetry of the scenario, and finish with the question of detecting genuine multipartite entanglement.

\subsection*{Case 1. Multipartite entanglement}
Let us then start by considering a state $\rho^{\rA:\rB:\rC}$ that is fully separable \ie
\be\label{fully sep}
\rho^{\rA:\rB:\rC}=\sum_\lambda p_\lambda \rho_\lambda^\rA\otimes\rho_\lambda^\rB\otimes\rho_\lambda^\rC,
\ee
where $p_\lambda$ defines a probability distribution.

\subsubsection*{Case 1A. Multipartite entanglement with one untrusted party}

We treat first the case where a single party, taken to be $\rA$, performs a set of untrusted measurements $\{M_{a|x}\}_{a,x}$ on her share of the state, providing the parties $\rB$ and $\rC$ with the assemblage
\ba\label{mult steering two parties}
\assem{\sigma}{a|x}{\rB\rC}&=&\Tr_\rA(M_{a|x}\otimes\openone_\rB\otimes\openone_\rC \rho^{\rA:\rB:\rC})\nonumber\\
&=&\sum_\lambda p(a|x,\lambda) \rho_\lambda^\rB\otimes\rho_\lambda^\rC.
\ea
where $p(a|x,\lambda)=p_\lambda\Tr(M_{a|x}\rho^\rA_\lambda)$. Notice first that the dependence of Bob and Charlie's assemblage on $a$ and $x$ comes only from the common pre-shared variable $\lambda$. This is a typical instance of an unsteerable assemblage, or, in other words, this is a local hidden state (LHS) model for the assemblage $\assem{\sigma}{a|x}{\rB\rC}$ \cite{WJD07}. Notice further that it is composed only by separable (unnormalised) states for $\rB$ and $\rC$. Thus, in this case, testing for the presence of multipartite entanglement reduces to testing if the assemblage $\assem{\sigma}{a|x}{\rB\rC}$ is steerable and separable at the same time.  This type of assemblages forms a set $\Sigma^{\rA:\rB:\rC}_{\rB\rC}$, given by
\begin{equation}\label{mult steering two parties set}
\Sigma^{\rA:\rB:\rC}_{\rB\rC} = \Big\{ \assem{\sigma}{a|x}{\rB\rC} \Big| \assem{\sigma}{a|x}{\rB\rC} = \textstyle{\sum_\mu} D(a|x,\mu) \rho_\mu^{\rB\rC}, \rho_\mu^{\rB\rC} \in \mathrm{SEP}  \Big\}
\end{equation}
where we have used the fact that any probability distribution $p(a|x,\lambda)$ can be written as a convex combination of deterministic ones $D(a|x,\mu)$, \ie $p_\lambda(a|x) = \sum_\mu q(\mu|\lambda)D(a|x,\mu)$, and denote the set of (unnormalised) separable quantum states by $\mathrm{SEP}$.

\subsubsection*{Case 1B. Multipartite entanglement with two untrusted parties} Now we consider that two of the parties, $\rA$ and $\rB$, have untrusted measuring devices. In this case they prepare an assemblage for Charlie, which is given by
\ba\label{mult steering single party}
\assem{\sigma}{ab|xy}{\rC}&=&\Tr_{\rA\rB}(M_{a|x}\otimes M_{b|y}\otimes\openone_\rC \rho^{\rA:\rB:\rC})\nonumber\\
&=&\sum_\lambda p_\lambda(ab|xy) \rho_\lambda^\rC.
\ea
Once again, since the only dependence of the assemblage on $a$, $b$, $x$ and $y$ is through $\lambda$ this assemblage is also unsteerable. Moreover, because the set of probability distributions (also called a behaviour) $p_\lambda(ab|xy)$ arises from local measurements on a separable state it must be local, \ie it can be written as $p_\lambda(ab|xy) = \sum_{\mu\nu} q(\mu\nu|\lambda)D(a|x,\mu)D(b|y,\nu)$ \cite{Bell review}. Therefore, the relevant set of assemblages $\Sigma^{\rA:\rB:\rC}_{\rC}$ is now given by
\begin{multline}\label{mult steering single party set}
\Sigma^{\rA:\rB:\rC}_{\rC} = \Big\{ \assem{\sigma}{ab|xy}{\rC} \Big| \assem{\sigma}{ab|xy}{\rC} = \textstyle{\sum_{\mu\nu}} D(a|x,\mu)D(b|y,\nu) \rho_{\mu\nu}^{\rC},\\ \rho_{\mu\nu}^{\rC} \geq 0  \Big\}.
\end{multline}

\subsection*{Case 2. Entanglement in a bipartition}

Let us now consider the case where the state $\rho^{ABC}$ is separable with respect to a single given bipartition. Choosing this partition to be $\rA: \rB\rC$ we now consider states of the form
\be\label{bisep}
\rho^{\rA:\rB\rC}=\sum_{\lambda}p_\lambda\rho^\rA_\lambda\otimes\rho^{\rB\rC}_\lambda.
\ee
Crucially, given the asymmetry of picking a bipartition as well as the asymmetry of picking the trusted party (or parties), we will see below that we have two inequivalent situations to consider, for both the cases of one or of two untrusted parties.

\subsubsection*{Case 2A. Entanglement in a bipartition with one untrusted party}
When only one of the parties performs uncharacterised measurements the asymmetry of \eqref{bisep} leads to two different situations: (i) when the lone party $\rA$ has the untrusted devices and (ii) when $\rB$ (or equivalently $\rC$) does.
In the first case Bob and Charlie's assemblage is given by
\ba\label{unsteerable A to BC}
\assem{\sigma}{a|x}{\rB\rC}&=&\Tr_\rA(M_{a|x}\otimes \openone_{\rB}\otimes\openone_{\rC}\rho^{\rA:\rB\rC})\nonumber\\
&=&\sum_\lambda p(a|x,\lambda)\rho^{\rB\rC}_\lambda,
\ea
Once again the dependence of Bob and Charlie's assemblage on $a$ and $x$ comes only from the common variable $\lambda$, and so this assemblage is unsteerable. In comparison to previously, the states distributed to Bob and Charlie are now arbitrary entangled states, and hence there is no additional structure that the decomposition imposes. The set $\Sigma^{\rA:\rB\rC}_{\rB\rC}$ defined by assemblages of the form \eqref{unsteerable A to BC} is therefore given by
\begin{equation}\label{e: sigma A:BC BC}
\Sigma^{\rA:\rB\rC}_{\rB\rC}  = \Big\{ \assem{\sigma}{a|x}{\rB\rC} \Big| \assem{\sigma}{a|x}{\rB\rC} = \textstyle{\sum_\mu} D(a|x,\mu) \rho_\mu^{\rB\rC}, \rho_\mu^{\rB\rC} \geq 0 \Big\}.
\end{equation}
In the second case, where Bob is the one not trusting his measurements, Alice and Charlie are left with the following assemblage:
\ba\label{unsteerable B to A}
\assem{\sigma}{b|y}{\rA\rC}&=&\Tr_\rB(\openone_{\rA}\otimes M_{b|y}\otimes \openone_{\rC}\rho^{\rA:\rB\rC})\nonumber\\
&=&\sum_\lambda p_\lambda \rho^{\rA}_\lambda\otimes\assem{\sigma}{b|y,\lambda}{\rC},
\ea
which is a fundamentally different situation. This assemblage now has two main features: (i) the only dependence on the variables $b$ and $y$ are due to Charlie's states. In other words, $\assem{\sigma}{b|y}{\rA\rC}$ is unsteerable from Bob to Alice but not necessarily from Bob to Charlie (note however that the no-signalling condition still holds from Bob to Charlie). This implies that if we trace out system C (or apply any quantum-to-classical map to it) the resulting assemblage for Alice alone will be unsteerable; (ii) it is composed by separable (unnormalised) states. The relevant set, $\Sigma^{\rA:\rB\rC}_{\rA\rC}$, is now given by
\begin{multline}\label{set uns B to A}
\Sigma^{\rA:\rB\rC}_{\rA\rC}  = \Big\{ \assem{\sigma}{b|y}{\rA\rC} \Big| \Tr_\rC  \assem{\sigma}{b|y}{\rA\rC} = \textstyle{\sum_\mu} D(b|y,\mu) \rho_{\mu}^{\rA},\\ \rho_{\mu}^{\rA} \geq 0, \assem{\sigma}{b|y}{\rA\rC} \in \mathrm{SEP}, \sum_{b}\assem{\sigma}{b|y}{\rA\rC} = \sum_{b}\assem{\sigma}{b|y'}{\rA\rC} \Big\}
\end{multline}

\subsubsection*{Case 2B. Entanglement in a bipartition with two untrusted parties} Again the asymmetry of the decomposition  \eqref{bisep} leads to two different situations: In one the untrusted measurements are at $\rA$ and $\rB$ (or similarly $\rA$ and $\rC$), whilst in the other they are at $\rB$ and $\rC$. In the first case the assemblage obtained is given by
\ba\label{sep A:BC unst A and B}
\assem{\sigma}{ab|xy}{\rC}&=&\Tr_{\rA\rB}(M_{a|x}\otimes M_{b|y}\otimes \openone_\rC \rho^{\rA:\rB\rC})\nonumber\\
&=&\sum_\lambda p(a|x,\lambda)\assem{\sigma}{b|y,\lambda}{\rC}.
\ea
This assemblage has only one main feature, that it may contain only steering from Bob to Charlie, and not from Alice to Charlie. It then defines the set $\Sigma^{\rA:\rB\rC}_{\rC}$ as
\begin{multline}\label{e:set A:BC C}
\Sigma^{\rA:\rB\rC}_{\rC} = \Big\{ \assem{\sigma}{ab|xy}{\rC} \Big| \assem{\sigma}{ab|xy}{\rC} = \textstyle{\sum_{\mu}} D(a|x,\mu) \assem{\sigma}{b|y,\mu}{\rC}, \\
\assem{\sigma}{b|y,\mu}{\rC} \geq 0, \sum_b\assem{\sigma}{b|y,\mu}{\rC} = \sum_b\assem{\sigma}{b|y',\mu}{\rC} \Big\}
\end{multline}

In the second case the resulting assemblage is given by
\ba\label{sep A:BC unst B and C}
\assem{\sigma}{bc|yz}{\rA}&=&\Tr_{\rB\rC}(\openone_\rA\otimes M_{b|y}\otimes M_{c|z} \rho^{\rA:\rB\rC})\nonumber\\
&=&\sum_\lambda p(bc|yz,\lambda)\rho^\rA_\lambda.
\ea
Here, there are two main features: (i) this assemblage is unsteerable; (ii) The behaviour $p(bc|yz,\lambda)$ arises from local measurements on a possibly entangled state $\rho_\lambda^{\rB\rC}$, it may contain nonlocal \textit{quantum} correlations \cite{Bell review}. The final set we define is therefore $\Sigma^{\rA:\rB\rC}_{\rA}$, given by
\begin{multline}\label{e:set quantum}
\Sigma^{\rA:\rB\rC}_{\rA} = \Big\{ \assem{\sigma}{bc|yz}{\rA} \Big| \assem{\sigma}{bc|yz}{\rA} = \textstyle{\sum_{\lambda}} p(bc|yz,\lambda) \assem{\sigma}{\lambda}{\rA}, \\ \assem{\sigma}{\lambda}{\rA} \geq 0, p(bc|yz,\lambda)\in \mathcal{Q}  \Big\}
\end{multline}
where we have denoted by $\mathcal{Q}$ the set of probability distributions which can arise from local measurements on quantum states.

\subsection*{Case 3. Genuine multipartite entanglement}
Let us now turn to the question of genuine multipartite entanglement (GME) detection. Genuine tripartite entangled states are the ones that can not be written as
\ba\label{bisep rho}
\rho^{\mathrm{bisep}}&=&\sum_\lambda p_\lambda^{\rA:\rB\rC}\rho^\rA_\lambda\otimes\rho^{\rB\rC}_\lambda+\sum_\lambda p_\lambda^{\rB:\rA\rC}\rho^\rB_\lambda\otimes\rho^{\rA\rC}_\lambda\nonumber\\&+&\sum_\lambda p_\lambda^{\rA\rB:\rC}\rho^{\rA\rB}_\lambda\otimes\rho^{C}_\lambda,
\ea
where $p_\lambda^{\rA:\rB\rC}$, $p_\lambda^{\rB:\rA\rC}$ and $ p_\lambda^{\rA\rB:\rC}$ are probability distributions. Our goal once again is to determine what constraints the form \eqref{bisep rho} imposes on the obtained assemblages, which will now follow straightforwardly given the analysis made before.

\subsubsection*{Case 3A. Genuine multipartite entanglement with one untrusted party}
When Alice is the one holding the untrusted devices, Bob and Charlie's assemblage is given by
\begin{widetext}
\ba\label{gen mult steering two parties}
\assem{\sigma}{a|x}{\rB\rC}&=&\Tr(M_{a|x}\otimes \openone_\rB\otimes\openone_\rC \rho^{\mathrm{bisep}})\nonumber\\
&=&\underbrace{\sum_\lambda p_\lambda^{\rA:\rB\rC}p(a|x,\lambda)\rho^{\rB\rC}_\lambda}_{\assem{\Gamma}{a|x}{\rA:\rB\rC} \in \Sigma^{\rA:\rB\rC}_{\rB\rC}}+\underbrace{\sum_\lambda p_\lambda^{\rB:\rA\rC}\rho^\rB_\lambda\otimes\assem{\sigma}{a|x,\lambda}{\rC}}_{\assem{\Gamma}{a|x}{\rB:\rA\rC} \in \Sigma^{\rB:\rA\rC}_{\rB\rC}}+\underbrace{\sum_\lambda p_\lambda^{\rA\rB:\rC}\assem{\sigma}{a|x\lambda}{\rB}\otimes\rho^{\rC}_\lambda}_{\assem{\Gamma}{a|x}{\rC:\rA\rB}\in \Sigma^{\rC:\rA\rB}_{\rB\rC}}.
\ea
\end{widetext}
The terms $\assem{\Gamma}{a|x}{\rA:\rB\rC}$ and $\assem{\Gamma}{a|x}{\rB:\rA\rC}$ can be seen as assemblages having the same structure as the assemblages \eqref{unsteerable A to BC} and  \eqref{unsteerable B to A} respectively, while the assemblage $\assem{\Gamma}{a|x}{\rC:\rA\rB}$ is identical to $\assem{\Gamma}{a|x}{\rB:\rA\rC}$, except that the role of Bob and Charlie is interchanged.

\subsubsection*{Case 3B. Genuine multipartite entanglement with two untrusted parties}
Consider now that Alice and Bob perform untrusted measurements, leading to:
\begin{widetext}
\ba\label{gen mult steering single party}
\assem{\sigma}{ab|xy}{\rC}&=&\Tr_{\rA\rB}(M_{a|x}\otimes M_{b|y}\otimes\openone_\rC \rho^{\mathrm{bisep}})\nonumber\\
&=&\underbrace{\sum_\lambda p_\lambda^{\rA:\rB\rC}p(a|x,\lambda)\assem{\sigma}{b|y,\lambda}{\rC}}_{\assem{\Pi}{ab|xy}{\rA:\rB\rC} \in \Sigma^{\rA:\rB\rC}_{\rC}}+\underbrace{\sum_\lambda p_\lambda^{\rB:\rA\rC}p(b|y,\lambda)\assem{\sigma}{a|x,\lambda}{\rC}}_{\assem{\Pi}{ab|xy}{\rB:\rA\rC}\in \Sigma^{\rB:\rA\rC}_{\rC}}+\underbrace{\sum_\lambda p_\lambda^{\rA\rB:\rC}p(ab|xy,\lambda)\rho^{\rC}_\lambda}_{\assem{\Pi}{ab|xy}{\rC:\rA\rB}\in \Sigma^{\rC:\rA\rB}_{\rC}}.
\ea
\end{widetext}
Again, the assemblages $\assem{\Pi}{ab|xy}{\rA:\rB\rC}$ and $\assem{\Pi}{ab|xy}{\rB:\rA\rC}$ are seen to have the same structure as \eqref{sep A:BC unst A and B}, while the assemblage $\assem{\Pi}{ab|xy}{\rC:\rA\rB}$ has the same structure as \eqref{sep A:BC unst B and C}.\\

\textbf{Supplementary note 2}\\

\textbf{SDP tests and semi-device-independent entanglement witnesses}\\
\label{a:sdp and ineqs}

 We have previously determined the constraints that each kind of entanglement imposes, and defined the corresponding sets of assemblages these constraints define. We now turn to the following practical question: given that we have observed a specific assemblage, can we test for a certain type of entanglement by checking whether or not the assemblage belongs to one of the previously defined sets?

Crucially, it turns out that all of the sets defined above are either specified solely in terms of positive semi-definite (PSD) constraints and linear matrix inequalities (LMIs), or can be approximated from the outside by a set with such a specification. Testing for membership inside such a set is an optimisation problem known as a semi-definite program (SDP), for which efficient numerical methods exist for the case of small systems, allowing for an answer to this question \cite{SDP}. Moreover, due to the theory of duality, the dual SDP provides us with a semi-device-independent witness that allow us to certify the presence of the different types of entanglement solely from the knowledge of the assemblage. This is similar to the ideas of entanglement witnesses and Bell inequalities in the standard and fully device-independent scenarios respectively.

\subsection{Deriving the SDP tests} \label{a: SDP tests}

In those cases where the the sets defined above are not specified solely in terms of PSD constraints and LMIs our strategy is to show that there exist suitable relaxations which are, that is to define bigger sets which are specified solely in terms of such constraints.

Working through in the order that they appeared, we shall consider each set in turn. The first set is $\Sigma^{\rA:\rB:\rC}_{\rB\rC}$, given in equation \eqref{mult steering two parties set}. It is the final constraint, $\rho_\mu^{\rB\rC} \in \mathrm{SEP}$, that does not have the desired form, as the set of (unnormalised) separable states has in general a complicated structure. The one case where the set in fact has a simple characterisation is if the dimensions satisfy $d_\rB d_\rC \leq 6$, in which case the set of separable states is exactly the set of states positive under partial transposition (PPT) \cite{HorPLA}. In this simple case we can rewrite $\rho_\mu^{\rB\rC} \in \mathrm{SEP}$ as $\left(\rho_\mu^{\rB\rC}\right)^{\rT_\rB} \geq 0$, where $\rT_\rB$ denotes the partial transposition with respect to system B. Since this operation is a linear map (on the state), this is now a PSD constraint, and the set is in fact in the desired form.

In all other dimensions, we can use the relaxation of the separable states to those that have a $k$-symmetric PPT extension \cite{DPS}. That is, we define the set $\mathrm{SYM}^{(k)}_{\rB\rC}$
\begin{multline}
\mathrm{SYM}^{(k)}_{\rB\rC} = \Big\{\rho^{\rB\rC} \Big| \rho^{\rB\rC} = \Tr_{\rB_2\cdots \rB_k} \rho^{\rB_1\cdots \rB_k\rC},\\
S_{ij} \rho^{\rB_1\cdots \rB_k\rC} S_{ij}^\dagger = \rho^{\rB_1\cdots \rB_k\rC} \quad\forall i\neq j,\\ \big(\rho^{\rB_1\cdots \rB_k\rC}\big)^{T_\rC} \geq 0, \rho^{\rB_1\cdots \rB_k\rC} \geq 0\Big\}
\end{multline}
where $S_{ij}$ is the swap operator between $\rB_i$ and $\rB_j$. This demands that $\rho^{\rB\rC}$ can be extended to a state with $k$ Bobs, which is symmetric under interchange and PPT, such that the reduced state of a single Bob and Charlie is the original state. Such sets are all specified in terms of PSD constraints and LMIs, and converge to the set of separable states as $k \to \infty$ \cite{DPS,DPSmulti}. For the case $k = 1$ it also reduces to the set of PPT state. We thus define the sequence of relaxations
\begin{multline}
\Sigma^{\rA:\rB:\rC(k)}_{\rB\rC} = \Big\{ \assem{\sigma}{a|x}{\rB\rC} \Big| \assem{\sigma}{a|x}{\rB\rC} = \textstyle{\sum_\mu} D(a|x,\mu) \rho_\mu^{\rB\rC}, \\
\rho_\mu^{\rB\rC} \in \mathrm{SYM}^{(k)}_{\rB\rC}\Big\},
\end{multline}
which now have the desired structure for each $k$.

Moving on, the set $\Sigma^{\rA:\rB:\rC}_{\rC}$ given in \eqref{mult steering single party set} already has the desired structure. This is also true for the set $\Sigma^{\rA:\rB\rC}_{\rB\rC}$ defined in \eqref{e: sigma A:BC BC}. The set $\Sigma^{\rA:\rB\rC}_{\rA\rC}$ \eqref{set uns B to A} contains the requirement that $\assem{\sigma}{b|y}{\rA\rC} \in \mathrm{SEP}$, which is dealt with in exactly the same way as above. Thus we define the relaxed set
\begin{multline}
\Sigma^{\rA:\rB\rC(k)}_{\rA\rC}  = \Big\{ \assem{\sigma}{b|y}{\rA\rC} \Big| \Tr_\rC  \assem{\sigma}{b|y}{\rA\rC} = \textstyle{\sum_\mu} D(b|y,\mu) \rho_{\mu}^{\rA},\\ \rho_{\mu}^{\rA} \geq 0, \assem{\sigma}{b|y}{\rA\rC} \in \mathrm{SYM}^{(k)}_{\rA\rC}, \sum_b \assem{\sigma}{b|y}{\rA\rC} = \sum_b \assem{\sigma}{b|y'}{\rA\rC} \Big\}
\end{multline}

The set $\Sigma^{\rA:\rB\rC}_{\rC}$ given in \eqref{e:set A:BC C} has the desired structure.

Finally, the set  $\Sigma^{\rC:\rA\rB}_{\rC}$ (which is has the same structure as \eqref{e:set quantum}) is less straightforward because of the constraint $p(ab|xy,\lambda) \in \mathcal{Q}$, \ie that the behaviours $p(ab|xy,\lambda)$ should have a quantum realisation. As in the nonlocality scenario of deciding if a behaviour has a quantum realisation, the exact answer to this problem is in general intractable. However, we can use the idea introduced in \cite{Pus13} (see also \cite{NavTorVer14}) to obtain a semi-definite relaxation. The basic idea is to apply the method of the NPA hierarchy \cite{NPA} only to the untrusted devices, whilst leaving Charlie quantum, using also the fact that the state is separable on the $\rA\rB:\rC$ partition. We thus relax to $\assem{\Pi}{ab|xy}{\rC:\rA\rB} \in \mathcal{Q}_\rC^{(k)}$, where $ \mathcal{Q}_\rC^{(k)}$ is defined by
\begin{multline}\label{e: Q^k def}
 \mathcal{Q}_\rC^{(k)} =  \Big\{ \assem{\sigma}{ab|xy}{\rC} \Big| \Gamma^{(k)}_\rC \geq 0, \left(\Gamma^{(k)}_\rC\right)^{\mathrm{T}_\rC} \geq 0 \\
  \Tr \big(G_j \Gamma^{(k)}_\rC\big) = \Tr \big(h_j \assem{\sigma}{ab|xy}{\rC}\big) \forall j \Big\}
\end{multline}
for some sets of operators $\{G_j^{(k)}\}$ and $\{h_j^{(k)}\}$, which encode the constraints that arise in the original NPA hierarchy \cite{NPA}, coming from (i) orthogonality of measurement outcomes, and (ii) commutivity of Alices and Bob measurements. Also, we can apply the idea from \cite{Pus13} and impose that the matrices $\Gamma_\rC^{(k)}$ are positive under partial transposition of Charlie (the trusted party), as a relaxation of the separability criteria across the bipartition. The main difference with the NPA approach is that whereas previously the elements of the matrix $\Gamma^{(k)}$ were complex numbers with certain ones equal to the nonlocal behaviour, now one should think of the elements as matrices (of the dimension of Charlie), with certain ones equal to members of the assemblage. 

Last, in order to be able to impose semidefinite constraints to the set $\Sigma^{\rC:\rA\rB}_{\rC}$ we need to constrain the number of terms in the summation in $\lambda$. We do this by noticing that any quantum behaviour can be written as a convex combination of extremal non-signalling behaviours $D^\rNS(bc|yz,\nu)$ \cite{Bell review}.

Given the above, set $\Sigma^{\rC:\rA\rB}_{\rA}$ is relaxed to
\begin{multline}
\Sigma^{\rC:\rA\rB(k)}_{\rC} = \Big\{ \assem{\sigma}{ab|xy}{\rC} \Big| \assem{\sigma}{ab|xy}{\rC} = \textstyle{\sum_{\nu}} D^\mathrm{NS}(ab|xy,\nu) \assem{\sigma}{\nu}{\rC}, \\ \assem{\sigma}{ab|xy}{\rC} \in \mathcal{Q}^{(k)}_\rC  \Big\}.
\end{multline}

Having found appropriate relaxations of all of the sets which we wish to consider, we can now straightforwardly write down an approximate optimisation problem in the form of an SDP that needs to be solved to check for the desired type of entanglement in each given scenario. Let us describe explicitly the approximate test that checks for the existence of a decomposition of the form \eqref{gen mult steering single party}, \ie that checks for genuine multipartite entanglement with two untrusted parties. We provide all the other SDP tests for the other decompositions described above in Supplementary Table \ref{t: SDPs}.

In \eqref{gen mult steering single party} we see that we have to find 3 assemblages, each contained in a different set, with each set either in a form directly usable, or for which we just gave an outer approximation above. Thus, by introducing the \textit{maximally mixed assemblage} $\assem{\mathrm{id}}{ab|xy}{\rC} = \frac{1}{m_\rA m_\rB}\openone_\rC/d_\rC$ we arrive at the following SDP test for genuine multipartite entanglement with 2 untrusted parties
\begin{align}\label{SDP gen mult}
\mathrm{max}\quad & p \nonumber \\
\mathrm{s.t.}\quad &  \assem{\Pi}{ab|xy}{\rA:\rB\rC} + \assem{\Pi}{ab|xy}{\rB:\rA\rC} + \assem{\Pi}{ab|xy}{\rC:\rA\rB} = \assem{\sigma}{ab|xy}{\obs} - p \,\assem{\mathrm{id}}{ab|xy}{\rC} \nonumber \\
&\assem{\Pi}{ab|xy}{\rA:\rB\rC} \in \Sigma^{\rA:\rB\rC}_{\rC}, \assem{\Pi}{ab|xy}{\rB:\rA\rC} \in \Sigma^{\rB:\rA\rC}_{\rC}, \\
&\quad\quad\quad \assem{\Pi}{ab|xy}{\rC:\rA\rB} \in \Sigma^{\rC:\rA\rB(k)}_{\rC} \nonumber
\end{align}
where $\assem{\sigma}{ab|xy}{\obs}$ is the observed assemblage of Charlie. Since $\assem{\mathrm{id}}{ab|xy}{\rC}$ is clearly contained in all 3 sets, being producible from the maximally mixed state, a sufficiently large negative $p$ will always be a solution, hence the SDP is strictly feasible. A strictly negative optimal solution $p^* < 0$ certifies that $\assem{\sigma}{ab|xy}{\obs}$ being measured does not have the desired decomposition, i.e. that the state is genuinely multipartite entangled. On the other hand an optimal value $p^* = 0$ indicates that a decomposition can be found. Note however that in this case, given the relaxation of the problem, one is not able to conclude anything regarding the separability of the state. One can take the parameter $k$ larger to obtain a better approximation to the original problem.

\subsection{Semi device-independent entanglement witnesses}\label{a: inequalities}
The dual of the  SDP  \eqref{SDP gen mult} is also readily written down \cite{SDP}, and is given by
\begin{align}\label{e:dual}
\min &\quad \Tr\sum_{abxy} \assem{F}{ab|xy}{} \assem{\sigma}{ab|xy}{\obs}\nonumber  \\
\text{s.t.} &\quad   \Tr\sum_{abxy} \assem{F}{ab|xy}{}\assem{\sigma}{ab|xy}{\rC} \geq 0\nonumber \\
&\quad\quad\quad\quad  \forall \,\, \assem{\sigma}{ab|xy}{\rC} \in  \Sigma^{\rA:\rB\rC}_{\rC} \cup  \Sigma^{\rB:\rA\rC}_{\rC} \cup  \Sigma^{\rC:\rA\rB(k)}_{\rC} \\
&\quad  \Tr\sum_{abxy} \assem{F}{ab|xy}{}\assem{\mathrm{id}}{ab|xy}{\rC} = 1 \nonumber
\end{align}
which is seen to constitute a witness for genuine multipartite entanglement. That is, the dual provides a set of operators $\{F_{ab|xy}\}_{abxy}$ such that the linear functional $\beta = \sum_{abxy}F_{ab|xy} \sigma_{ab|xy}$ is greater than zero for all assemblages which arise from measurements on a bi-separable state. An observed value $\beta^\obs < 0$ thus provides a witness which certifies the genuine multipartite entanglement of the state in a semi-device-independent manner. The final condition, $\Tr\sum_{abxy} \assem{F}{ab|xy}{}\assem{\mathrm{id}}{ab|xy}{\rC} = 1$ is a convention, which simply defines an overall scale for the witness.

More generally, the dual of each SDP in Supplementary Table \ref{t: SDPs} provides witness operators $\{\assem{F}{a|x}{}\}_{ax}$, for the case of Alice untrusted, or $\{\assem{F}{ab|xy}{}\}_{abxy}$, for the case of Alice and Bob untrusted (or a permutation of the parties) which constitute a semi-device-independent entanglement witnesses of the form
\begin{align}
	&\Tr \sum_{ax} \assem {F}{a|x}{}\assem{\sigma}{a|x}{} \geq 0\quad \forall \assem{\sigma}{a|x}{} \in \Sigma\nonumber\\
&\Tr \sum_{abxy} \assem {F}{ab|xy}{}\assem{\sigma}{ab|xy}{} \geq 0 \quad \forall \assem{\sigma}{ab|xy}{} \in \Sigma'
\end{align}
with corresponding violations $\beta^\obs = \Tr \sum_{ax} \assem {F}{a|x}{}\assem{\sigma}{a|x}{\obs} < 0$ or $\beta^\obs = \Tr \sum_{ax} \assem {F}{ab|xy}{}\assem{\sigma}{ab|xy}{\obs} < 0$ respectively, where $\Sigma$ and $\Sigma'$ are sets, or union of sets (depending upon the type of entanglement one is checking for), as defined above.

Finally, we note that it is possible to put these witnesses into two more friendly forms in the case of binary measurement outcomes, so-called \emph{observable} and \emph{coefficient} forms. Starting with the observable form, we use the definition of the observed assemblage, and introduce the observables $A_x = M_{0|x} - M_{1|x}$ to arrive at
\begin{align}
	&\Tr \sum_{ax}  \assem{F}{a|x}{}\assem{\sigma}{a|x}{\obs} = \Tr\sum_{ax}  M_{a|x} \otimes \assem{F}{a|x}{} \rho \nonumber \\
	&= \tfrac{1}{2}\Tr\sum_{ax}  (\openone_\rA + (-1)^a A_x) \otimes \assem{F}{a|x}{} \rho \nonumber \\
	&= \Tr\big( \openone_\rA \otimes \tfrac{1}{2}\sum_{ax}\assem{F}{a|x}{} + \sum_{x} A_x  \otimes \tfrac{1}{2}\sum_{a} (-1)^a \assem{F}{a|x}{} \big)\rho \nonumber \\
	&= \Tr\big( \openone_\rA \otimes J_\emptyset + \sum_{x} A_x \otimes J_x \big)\rho
\end{align}
for the case of one untrusted party, where we have defined the observables $J_\emptyset = \tfrac{1}{2}\sum_{ax}\assem{F}{a|x}{}$ and $J_x = \tfrac{1}{2}\sum_{a} (-1)^a \assem{F}{ab|xy}{}$ for Bob and Charlie. For the coefficient form we further expand these matrices in the (complete) basis of Pauli operators, namely,
\begin{align}
J_\emptyset &= \sum_{yz}f_{0yz}\sigma^y \otimes \sigma^z,& J_x = \sum_{yz}f_{xyz}\sigma^y \otimes \sigma^z
\end{align}
where $\sigma^0 = \openone$, $\sigma^1 = X$, $\sigma^2 = Y$ and $\sigma^3 = Z$ and denoting $A_0 = \openone$ then we arrive at the compact form
\begin{align}
	&\Tr \sum_{ax}  \assem{F}{a|x}{}\assem{\sigma}{a|x}{\obs} =\Tr \left(\sum_{xyz} (f_{xyz} A_x \otimes \sigma^y\otimes \sigma^z)\rho\right)
\end{align}

For the case of two untrusted parties, an analogous but longer calculation gives for the observable form
\begin{multline}
	\Tr \sum_{abxy}  \assem{F}{ab|xy}{}\assem{\sigma}{ab|xy}{\obs} \\
	= \Tr\big( \openone_\rA \otimes \openone_\rB \otimes K_\emptyset + \sum_{x} A_x \otimes \openone_\rB \otimes K_x \\
	\quad + \sum_y \openone_\rA \otimes B_y \otimes K'_y + \sum_{xy} A_x\otimes B_y \otimes K_{xy} \big)\rho
\end{multline}
where we have introduced the observables $B_y$ for Bob, as well as the observables $K_\emptyset = \tfrac{1}{4}\sum_{abxy}\assem{F}{ab|xy}{}$, $K_x = \tfrac{1}{4}\sum_{aby} (-1)^a \assem{F}{ab|xy}{}$, $K'_y = \tfrac{1}{4}\sum_{abx}(-1)^b \assem{F}{ab|xy}{}$ and $K_{xy} = \tfrac{1}{4}\sum_{ab} (-1)^{a+b} \assem{F}{ab|xy}{}$ for the trusted party Charlie, that need to be measured in the corresponding configurations, given above. For the coefficient form, again by expanding these matrices in the complete basis of Pauli matrices, 
\begin{align}
K_\emptyset &= \sum_{z}f_{00z}\sigma^z,& K_x = \sum_{z}f_{x0z} \sigma^z \nonumber \\
K'_y &= \sum_{z}f_{0yz}\sigma^z,& K_{xy} = \sum_{z}f_{xyz} \sigma^z
\end{align}
and by denoting $B_0 = \openone$, we obtain the analogous compact form as previously, 
\begin{align}
\Tr \sum_{abxy}  \assem{F}{ab|xy}{}\assem{\sigma}{ab|xy}{\obs} = \Tr \left(\sum_{xyz} (f_{xyz} A_x \otimes B_y\otimes \sigma^z)\rho\right)
\end{align}
\\

\textbf{Supplementary note 3}\\

\textbf{Generalisation to more parties}\label{a: generalisation}
\\

We have presented our main results in the tripartite case. Notice however that the same procedure can readily be followed to derive SDPs to test the presence of different kinds of entanglement for general $N$-partite systems.

First of all one specifies the scenarios by fixing (i) a particular type of entanglement and (ii) the pattern of trusted and untrusted parties. The entanglement can be chosen arbitrarily, for example one may ask that the state is not fully separable, be separable across a given number of fixed bipartitions, or be a convex combination of states separable over a given number of partitions (but not necessarily fixed). The patten may also be chosen arbitrarily, ranging from all but one party trusted, to all but one untrusted.

Given the specification, one then enumerates the list of properties which the corresponding assemblages have. These properties will fall into two classes - those which impose constraints which are directly applicable, \ie are in the form of PSD constraints and LMIs, and those which are not. As in the tripartite case, the objective is then to relax the non-directly applicable constraints to find an approximate SDP test.

The main difficultly in our approach is that as the number of parties increases, and the local dimension of the Hilbert space, we expect that the difficulty of the problem will grow to the point where current numerical techniques are unable to solve efficiently the tests. For example, one class of constraints that will arise is that multipartite assemblages will need to have quantum realisations. In principle such a constraint can still be imposed by applying the NPA hierarchy \cite{NPA} to the untrusted devices, however in the multipartite setting this soon becomes intractable. Alternatively, one may have constraints that a multipartite quantum state is separable. One can again relax this using the generalisation of the $k$-shareability condition \cite{DPSmulti}.

In summary, the approach presented here is most suitable to scenarios involving relatively small numbers of parties, where it provides powerful tests for multipartite entanglement (and explicitly provides witnesses in each case). This is however expected as this is also the case in standard entanglement detection techniques \cite{ent detection} (due to the increase of the Hilbert space dimension)  and in the fully device-independent approach \cite{Bell review} (due to the number of the space of local probability distributions).\\

\textbf{Supplementary note 4}\\

\textbf{Examples: GHZ and W states}\\

In order to demonstrate the usefulness of our previous characterisation we apply the above SDP to two exemplary genuine multipartite states, namely the GHZ and W states. More specifically we are interested in how much white noise can be added to these states until our method fails to detect either entanglement or GME, \ie we want find the minimum $w$, denoted by $w^*$, allowing us to detect either entanglement or GME in the states
\ba
&\rho_{\rGHZ}&=w\ket{\rGHZ}\bra{\rGHZ}+(1-w)\openone/8;\nonumber\\
&\rho_{W}&=w\ket{W}\bra{W}+(1-w)\openone/8,
\ea
where $\ket{\rGHZ}=(\ket{000}+\ket{111})/{\sqrt{2}}$ and  $\ket{W}=(\ket{001}+\ket{010}+\ket{100})/{\sqrt{3}}$.
Supplementary Table \ref{examples} gives a summary of the results, in terms of the numbers provided by our methods and a comparison to what was known regarding entanglement witnesses and Bell inequalities. All results were obtained using {\sc cvx} \cite{SDP} for {\sc Matlab} to solve the SDP, and the optimisation toolbox to numerically search for the best choices of measurements for Alice (and Bob). Since such a search over measurements choices provides no guarantee that the global optimum is obtained, all results constitute upper bounds. However, all of our numerical evidence suggests that the values obtained cannot be improved.

As we can see the values of $w^*$ lies in between the bound for entanglement, where the largest number of assumptions are made, and the bound from nonlocality, where no assumptions are made. Furthermore, as one would expect, stronger bounds are possible with 2 parties trust their devices compared to the case of only 1.

We end by presenting the steering witnesses we obtain in the above for the GHZ and W states that certify genuine tripartite entanglement in a semi-device-independent fashion.

Starting with the GHZ state and the case of two untrusted parties (and three measurements), the optimal witness is
\begin{multline}\label{eq:GHZtwountrusted}
1 - \alpha \langle A_2 B_2 \rangle - \alpha \langle A_2 Z \rangle - \alpha \langle B_2 Z \rangle  - \beta \langle A_0 B_0 X \rangle  \\
+ \beta \langle A_0 B_1 Y \rangle + \beta \langle A_1 B_0 Y \rangle + \beta \langle A_1 B_1 X \rangle \geq 0
\end{multline}
where $\alpha = 0.1831$ and $\beta = 0.2582$, and the pure GHZ state achieves a violation $-0.5821\ngeq 0$. For the case of the GHZ state and only a single untrusted party, the witness is
\begin{multline}\label{eq:GHZoneuntrusted}
1
+ 0.1547 \expect{Z_\rB Z_\rC}
- \tfrac{1}{3}\big(\expect{A_2 Z_\rB} +\expect{A_2 Z_\rC} + \expect{A_0 X_\rB X_\rC} \\- \expect{A_0 Y_\rB Y_\rC} - \expect{A_1 X_\rB Y_\rC} - \expect{A_1 Y_\rB X_\rC}\big) \geq 0
\end{multline}
with the pure GHZ state now achieving a violation of $-0.8453\ngeq 0$. Interestingly, we note first that the structure of both witnesses is the same, the only difference being in the coefficients. Furthermore the only terms which appear are those which arise from the stabiliser relations of the GHZ state.

Moving on to the W state, for two untrusted parties the optimal witness is
\begin{multline}\label{eq:Wtwountrusted}
1
+ 0.2517\big(\expect{A_2} + \expect{B_2}\big)
+ 0.3520 \expect{Z}
- 0.1112\big(\expect{A_0 X} \\+ \expect{A_1 Y} + \expect{B_0 X} + \expect{B_1 Y}\big)
+ 0.1296\big(\expect{A_2Z} + \expect{B_2Z}\big)\\
-0.1943\big(\expect{A_0B_0} + \expect{A_1B_1}\big)
+ 0.2277 \expect{A_2B_2}\\
 -0.1590\big(\expect{A_0B_0Z} + \expect{A_1B_1Z}\big)
 + 0.2228 \expect{A_2B_2Z}\\
-0.2298\big(\expect{A_0B_2X} +\expect{A_1B_2Y} + \expect{A_2B_0X} + \expect{A_2B_1Y}\big) \geq 0
\end{multline}
and the pure W state obtains the violation $-0.4803 \ngeq 0$. For one untrusted party the witness is
\begin{multline}\label{eq:Woneuntrusted}
1
+ 0.4405\big(\expect{Z_\rB} + \expect{Z_\rC}\big)
- 0.0037\expect{Z_\rB Z_\rC}
- 0.1570\big(\expect{X_\rB X_\rC} \\+ \expect{Y_\rB Y_\rC} + \expect{A_2 X_\rB X_\rC} + \expect{A_2 Y_\rB Y_\rC} \big)
+ 0.2424\big(\expect{A_2}\\+\expect{A_2 Z_\rB Z_\rC}\big)
+ 0.1848\big(\expect{A_2 Z_\rB} + \expect{A_2 Z_\rC}\big)
-0.2533\big(\expect{A_0 X_B} \\+ \expect{A_0 X_\rC} + \expect{A_1 Y_\rB} + \expect{A_1 Y_\rC} + \expect{A_0 X_\rB Z_\rC} + \expect{A_0 Z_\rB X_\rC} \\+ \expect{A_1 Y_\rB Z_\rC} + \expect{A_1 Z_\rB Y_\rC}\big) \geq 0
\end{multline}
with he pure W state achieving the violation $-0.7594 \ngeq 0$. Again, we note that structurally the witnesses are the same in the case of one and two untrusted parties.
\\

\textbf{Supplementary note 5}\\

\textbf{Experimental inequalities}\\

In this section we give the semi-device independent entanglement witnesses that were used to optimally certify the presence of genuine multipartite entanglement of the GHZ state. Starting with the case of one untrusted party, in coefficient form the inequality is given by
\begin{align}
f_{xy0} = 
\left(\begin{array}{rrrr}
 1.0000& -0.0183&  0.1079&  0.0130 \\ 
 0.1522&  0.1518& -0.0734&  0.0251 \\ 
-0.2870& -0.1125& -0.1229&  0.0189 \\ 
-0.0658&  0.0095& -0.0151& -0.2142
\end{array} \right), \nonumber \\
f_{xy1} = 
\left(\begin{array}{rrrr}
    0.2096&  0.0941&  0.1565&  0.0077 \\
    0.0633&  0.3006& -0.0195&  0.0132 \\
   -0.0388& -0.0133& -0.3040& -0.0033 \\
    0.0614&  0.0040& -0.0389& -0.0387 \\
\end{array} \right),\nonumber \\
f_{xy2} = 
\left(\begin{array}{rrrr}
   -0.1487& -0.1841&  0.0862& -0.0061 \\
   -0.1839&  0.0323&  0.2801&  0.0028 \\
    0.1036&  0.2920& -0.0423&  0.0071 \\
    0.0066&  0.0683&  0.0277&  0.0242 \\
\end{array} \right), \\
f_{xy3} = 
\left(\begin{array}{rrrr}
    0.0189&  0.0114&  0.0363& -0.2187 \\
   -0.0259& -0.0144&  0.0255& -0.0204\\
   -0.0714& -0.0027&  0.0099&  0.0505\\
   -0.0189&  0.1263& -0.2037&  0.0145\\
   \end{array}\right), \nonumber
\end{align}
where $x$ labels the rows and $y$ the columns. For two untrusted parties the inequality is given by
\begin{align}
f_{xy0} = 
\left(\begin{array}{rrrr}
    1.0000&    0.2375&   -0.2613&    0.0023\\
    0.0026&   -0.0041&   -0.0055&    0.0241\\
   -0.0039&   -0.0055&   -0.0114&    0.0291\\
   -0.0012&    0.0014&    0.0032&   -0.1515\\
\end{array} \right), \nonumber \\
f_{xy1} = 
\left(\begin{array}{rrrr}   
    0.0041&   -0.0033&   -0.0022&    0.0275\\
    0.1673&    0.2676&   -0.0317&    0.0044\\
    0.1710&   -0.0015&   -0.2737&   -0.0072\\
    0.0009&    0.0059&    0.0004&    0.0001\\
\end{array} \right), \nonumber \\
f_{xy2} = 
\left(\begin{array}{rrrr}    
   -0.0063&   -0.0091&    0.0053&    0.0106\\
   -0.1669&   -0.0004&    0.2651&    0.0073\\
    0.1675&    0.2569&   -0.0307&    0.0036\\
    0.0019&    0.0161&    0.0127&    0.0004\\
\end{array} \right), \\
f_{xy3} = 
\left(\begin{array}{rrrr}    
   -0.0013&    0.0054&    0.0011&   -0.1523\\
    0.0024&    0.0089&    0.0053&    0.0002\\
    0.0023&    0.0053&    0.0047&    0.0001\\
   -0.1527&   -0.0050&   -0.0085&   -0.0014\\
\end{array} \right). \nonumber 
\end{align}

\end{appendix}

\end{document}